\documentclass[11pt]{article} 
\usepackage{latexsym}

\def\hybrid{\topmargin -25pt    \oddsidemargin 0in 
	\headheight 0pt \headsep 0pt
	\textwidth 6.5in       
	\textheight 9.5in       
	\marginparwidth .875in
	\parskip 5pt plus 1pt   \jot = 1.5ex}
\hybrid
\def\cQ{{\cal Q}}
\def\cG{{\cal G}}
\def\cL{{\cal L}}

\def\cM{{\cal M}}
\def\cH{{\cal H}}
\def\ket#1{|{#1}\rangle}
\def\noi{\noindent}

\def\baselinestretch{1.2}

\catcode`\@=11

\def\marginnote#1{}
\def\draftlabel#1{{\@bsphack\if@filesw {\let\thepage\relax
   \xdef\@gtempa{\write\@auxout{\string
      \newlabel{#1}{{\@currentlabel}{\thepage}}}}}\@gtempa
   \if@nobreak \ifvmode\nobreak\fi\fi\fi\@esphack}
	\gdef\@eqnlabel{#1}}
\def\@eqnlabel{}
\def\@vacuum{}
\def\draftmarginnote#1{\marginpar{\raggedright\scriptsize\tt#1}}

\def\draft{\oddsidemargin -.2truein
	\def\@oddfoot{\sl preliminary draft \hfil
	\rm\thepage\hfil\sl\today\quad\militarytime}
	\let\@evenfoot\@oddfoot \overfullrule 3pt
	\let\label=\draftlabel
	\let\marginnote=\draftmarginnote
   \def\@eqnnum{(\theequation)\rlap{\kern\marginparsep\tt\@eqnlabel}%
\global\let\@eqnlabel\@vacuum}  }

\def\preprint{\twocolumn\sloppy\flushbottom\parindent 2em
	\leftmargini 2em\leftmarginv .5em\leftmarginvi .5em
	\oddsidemargin -.5in    \evensidemargin -.5in
	\columnsep .4in \footheight 0pt
	\textwidth 10.in        \topmargin  -.4in
	\headheight 12pt \topskip .4in
	\textheight 6.9in \footskip 0pt
	\def\@oddhead{\thepage\hfil\addtocounter{page}{1}\thepage}
	\let\@evenhead\@oddhead \def\@oddfoot{} \def\@evenfoot{} }

\def\numberbysection{\@addtoreset{equation}{section}
	\def\theequation{\thesection.\arabic{equation}}}

\def\underline#1{\relax\ifmmode\@@underline#1\else
	$\@@underline{\hbox{#1}}$\relax\fi}

\def\titlepage{\@restonecolfalse\if@twocolumn\@restonecoltrue
\onecolumn
     \else \newpage \fi \thispagestyle{empty}\c@page\z@
	\def\thefootnote{\fnsymbol{footnote}} }

\def\endtitlepage{\if@restonecol\twocolumn \else \newpage \fi
	\def\thefootnote{\arabic{footnote}}
	\setcounter{footnote}{0}}  

\catcode`@=12
\relax

\def\figcap{\section*{Figure Captions\markboth
	{FIGURECAPTIONS}{FIGURECAPTIONS}}\list
	{Figure \arabic{enumi}:\hfill}{\settowidth\labelwidth{Figure
999:}
	\leftmargin\labelwidth
	\advance\leftmargin\labelsep\usecounter{enumi}}}
 \relax
\def\tablecap{\section*{Table Captions\markboth
	{TABLECAPTIONS}{TABLECAPTIONS}}\list
	{Table \arabic{enumi}:\hfill}{\settowidth\labelwidth{Table
999:}
	\leftmargin\labelwidth
	\advance\leftmargin\labelsep\usecounter{enumi}}}
 \relax
\def\reflist{\section*{References\markboth
	{REFLIST}{REFLIST}}\list
	{[\arabic{enumi}]\hfill}{\settowidth\labelwidth{[999]}
	\leftmargin\labelwidth
	\advance\leftmargin\labelsep\usecounter{enumi}}}
 \relax
\makeatletter
\newcounter{pubctr}
\def\publist{\@ifnextchar[{\@publist}{\@@publist}}
\def\@publist[#1]{\list
	{[\arabic{pubctr}]\hfill}{\settowidth\labelwidth{[999]}
	\leftmargin\labelwidth
	\advance\leftmargin\labelsep
	\@nmbrlisttrue\def\@listctr{pubctr}
	\setcounter{pubctr}{#1}\addtocounter{pubctr}{-1}}}
\def\@@publist{\list
	{[\arabic{pubctr}]\hfill}{\settowidth\labelwidth{[999]}
	\leftmargin\labelwidth
	\advance\leftmargin\labelsep
	\@nmbrlisttrue\def\@listctr{pubctr}}}
 \relax
\makeatother
\newskip\humongous \humongous=0pt plus 1000pt minus 1000pt

\newif\ifdtup

\font\Scbig=cmss10 scaled\magstep1
\font\Scscr=cmss8 scaled\magstep1
\font\Scscrscr=cmss8
\newfam\Scfam
\textfont\Scfam=\Scbig
\scriptfont\Scfam=\Scscr
\scriptscriptfont\Scfam=\Scscrscr
\def\Sc{\fam\Scfam}

\relax
\def\lvm{\leavevmode\hbox to\parindent{\hfill}}

\def\BE{\begin{equation}}
\def\EE{\end{equation}}
\def\BA{\begin{eqnarray}}
\def\EA{\end{eqnarray}}

\def\D{\Delta}

\def\a{\alpha}

\def\tt{\bar\tau}

\def\lvm{\leavevmode\hbox to\parindent{\hfill}}
\def\bar{\overline}
\def\req#1{(\ref{#1})}

\def\L{\left}
\def\R{\right}

\def\BE{\begin{equation}}
\def\EE{\end{equation} \vskip 0.30\baselineskip}
\def\BA{\begin{array}}
\def\EA{\end{array}}

\def\noi{\noindent}

\def\frac#1#2{{\textstyle{{#1}\over{#2}}}}

\def\Kr#1{\delta_{{#1},0}}
\def\ket#1{|{#1}\rangle}

\def\cA{{\cal A}}
\def\cG{{\cal G}}
\def\cH{{\cal H}}

\def\cL{{\cal L}}

\def\cQ{{\cal Q}}

\def\open#1{\mbox{{\bf{#1}}}}

\def\oZ{{\open Z}}

\def\ctop{{\Sc c}}
\def\htop{{\Sc h}}

\def\a{\alpha}

\def\svec{singular vector}

\def\Qz{\cQ_0}
\def\Gz{\cG_0}
\def\Qn{$\Qz$}
\def\Gn{$\Gz$}
\def\kc{{\ket{\chi}}}
\def\kcc#1#2#3{{\kc_{#1}^{({#2}){#3}}}}

\newif\ifold \oldtrue \def\new{\oldfalse}
\let\ssection=\section
\def\section{\setcounter{equation}{0}\ssection}

\begin{document}
\renewcommand{\theequation}{\thesection.\arabic{equation}}
\newcommand{\beq}{\begin{equation}}
\newcommand{\eeq}[1]{\label{#1}\end{equation}}
\newcommand{\bea}{\begin{eqnarray}}
\newcommand{\eea}{\end{eqnarray}}
\newcommand{\eer}[1]{\label{#1}\end{eqnarray}}
\begin{titlepage}
\begin{center}

\hfill IMAFF-FM-99/09\\
\hfill hep-th/9910121
\vskip .4in

{\large \bf 
A Note concerning Subsingular Vectors and Embedding Diagrams of the 
N=2 Superconformal Algebras}                        
\vskip .4in

{\bf Beatriz Gato-Rivera}$^{a,b}$ \\
\vskip .3in

${ }^a${\em Instituto de Matem\'aticas y F\'\i sica Fundamental, CSIC \\
Serrano 123, Madrid 28006, Spain} \footnote{e-mail address:
bgato@pinar1.csic.es}\\

${ }^b${\em NIKHEF-H, Kruislaan 409, NL-1098 SJ Amsterdam, The Netherlands}\\

\vskip 0.3in

\end{center}

\begin{center} {\bf ABSTRACT } \end{center}
\begin{quotation}
Subsingular vectors of the N=2 superconformal algebras were discovered, and 
examples given, in 1996. Shortly afterwards Semikhatov and Tipunin claimed to 
have obtained a complete classification of the N=2 subsingular vectors in the 
paper `The Structure of Verma Modules over the N=2 Superconformal algebra', 
hep-th/9704111, published in CMP 195 (1998) 129. Surprisingly, the only 
explicit examples of N=2 subsingular vectors known at that time did not fit
into their classification. All the results presented in that paper, including
the classification of subsingular vectors, were based on the following 
assumptions: i) The authors claimed that there are only two different types of 
submodules in N=2 Verma modules, overlooking from the very beginning 
indecomposable `no-label' singular vectors, that had been discovered a few
months before, and clearly do not fit into their two types of submodules, and 
ii) The authors claimed to have constructed `non-conventional' singular vectors 
with the property of generating the two types of submodules maximally, i.e.
with no subsingular vectors left outside. In this note we prove that both
assumptions are incorrect. These facts also affect profoundly the results 
presented in several other publications, especially the papers: `On the 
Equivalence of Affine sl(2) and N=2 ....', by Semikhatov, hep-th/9702074,
`Embedding Diagrams of N=2 Verma Modules ....', by Semikhatov and Sirota, 
hep-th/9712102, and `All Singular Vectors of the N=2 ....', by Semikhatov and 
Tipunin, hep-th/9604176 (last revised version in September 98).

\end{quotation}
\vskip 1.2cm

October 1999
\end{titlepage}

\def\baselinestretch{1.2}
\baselineskip 16 pt
\section{Introduction and Notation}\lvm

The Topological N=2 superconformal algebra was deduced in 1990
as the symmetry algebra of two-dimensional topological conformal field 
theory (TCFT) \cite{DVV}. It was the
last N=2 superconformal algebra to be discovered and in fact can be obtained
from the Neveu-Schwarz N=2 algebra by modifying 
the stress-energy tensor by adding the derivative of the U(1) current, a
procedure known as {\it topological twist} \cite{EY}\cite{W-top}. It reads
\BE\new\BA{lclclcl}
\L[\cL_m,\cL_n\R]&=&(m-n)\cL_{m+n}\,,&\qquad&[\cH_m,\cH_n]&=
&{\ctop\over3}m\Kr{m+n}\,,\\
\L[\cL_m,\cG_n\R]&=&(m-n)\cG_{m+n}\,,&\qquad&[\cH_m,\cG_n]&=&\cG_{m+n}\,,
\\
\L[\cL_m,\cQ_n\R]&=&-n\cQ_{m+n}\,,&\qquad&[\cH_m,\cQ_n]&=&-\cQ_{m+n}\,,\\
\L[\cL_m,\cH_n\R]&=&\multicolumn{5}{l}{-n\cH_{m+n}+{\ctop\over6}(m^2+m)
\Kr{m+n}\,,}\\
\L\{\cG_m,\cQ_n\R\}&=&\multicolumn{5}{l}{2\cL_{m+n}-2n\cH_{m+n}+
{\ctop\over3}(m^2+m)\Kr{m+n}\,,}\EA\qquad m,~n\in\oZ\,.\label{topalgebra}
\EE
\noi
where $\cL_m$ and $\cH_m$ are the bosonic generators corresponding to the
stress-energy tensor (Virasoro generators) and the U(1) current,
respectively, and $\cG_m$ and $\cQ_m$ are the spin-2 and spin-1 fermionic 
generators, the latter being the modes of the BRST-current.
The eigenvalues of the bosonic zero modes $(\cL_0,\,\cH_0)$ correspond to
the conformal weight and the U(1) charge of the states. 
In a Verma module these eigenvalues split
conveniently as $(\D+l,\,\htop+q)$ for secondary states, where $l$
and $q$ are the {\it level} and the {\it relative charge} of the state and
$(\D,\,\htop)$ are the conformal weight and the charge of
the primary state on which the secondary is built.
The `topological' central charge $\ctop$ is the central charge 
corresponding to the Neveu-Schwarz N=2 algebra. The determinant 
formula for this algebra has remained unpublished until very recently 
\cite{DB4}, although the `chiral determinant formula', which applies to 
chiral Verma modules was published in 1997 \cite{BJI7}. 

Due to the existence of the fermionic zero modes \Gn\ and \Qn\  
this algebra has two sectors: the $\cG$-sector (states annihilated 
by \Gn ) and the $\cQ$-sector (BRST-invariant states annihilated by 
\Qn ), in analogy 
with the $(+)$-sector and the $(-)$-sector of the Ramond N=2 
algebra, due to the fermionic zero modes $G_0^+$ and $G_0^-$. However,
the two sectors do not provide the complete description --
and this is true for the Ramond N=2 algebra as well --
since there are also states which do not belong to any of the sectors
\cite{BJI6}\cite{DB2}\cite{DB4}. That is, not all Verma modules and
submodules decompose into the two sectors, but there are also 
indecomposable states, in particular indecomposable \svec s.
To see this one only needs to inspect the anticommutator of the
fermionic zero modes $\{\Gz , \Qz \}=2\cL_0$ acting on a given state $\kc$. 
If the conformal weight of $\kc$ is different from zero; i.e.
$\cL_0 \kc = (\D+l) \kc \neq 0$, then $\kc$ can be decomposed into a
state $\kc^G$ annihilated by \Gn , but not by \Qn , that we refer as
\Gn-closed and a state $\kc^Q$ annihilated by \Qn , 
but not by \Gn , that we refer as \Qn-closed:
\BE \kc = {1\over 2\D} \Qz \Gz \kc + {1\over 2\D} \Gz \Qz \kc = 
\kc^Q + \kc^G    \,. \EE   

If the conformal weight of $\kc$ is zero, however, one only obtains
$(\Gz \Qz + \Qz \Gz)\kc=0$, which is satisfied in four different ways: 
i) The state is \Gn-closed, $\kc=\kc^G$, and $\Gz \Qz \kc^G =0$,
ii) The state is \Qn-closed, $\kc=\kc^Q$, and $\Qz \Gz \kc^Q =0$,
iii) The state is chiral, $\kc=\kc^{G,Q}$, annihilated by both \Gn\ and \Qn ,
and iv) The state is indecomposable `no-label', $\kc=\kc$, not annihilated
by any of the fermionic zero modes. 
   
In what follows we will use the standard definition of 
highest weight vectors and \svec s for conformal 
algebras, i.e. they are the states with {\it lowest} conformal weight
({\it lowest} energy) in the Verma modules and in the null submodules, 
respectively, and therefore are annihilated by all the positive
modes of the generators of the algebra (the lowering operators); i.e. 
${\ } \cL_{n \geq 1} \kc =  \cH_{n \geq 1} \kc =  {\cG}_{n \geq 1} \kc
=  {\cQ}_{n \geq 1} \kc = 0 {\ }$. Hence these annihilation conditions will be 
referred to as the conventional, standard highest weight (h.w.) conditions. 
Singular vectors that are not generated by acting with the algebra on 
other \svec s are called {\it primitive}, otherwise they are called
{\it secondary} \svec s.

Subsingular vectors are also null but they do not satisfy the h.w.
conditions, becoming singular, that is annihilated by all the 
positive generators, in the quotient of the Verma module by a
submodule, however. As a consequence they are 
located {\it outside} that particular submodule (otherwise they would
disappear after taking the quotient), although descending to it 
necessarily by the action of the lowering operators (so that they
descend to `nothing' once the submodule is set to zero). 
This implies that
{\it the \svec s cannot reach the subsingular vectors } going upwards by 
the action of the negative, rising operators, whereas {\it the subsingular
vectors can reach the \svec s } going downwards by the action of the
positive, lowering operators.   

Subsingular vectors for the N=2 algebras were discovered in 1996 in
ref. \cite{BJI5} and the first examples for the case of the Topological N=2
algebra were published in January 1997 in ref. \cite{BJI6}, together with
the classification of all possible types of singular vectors taking into
account the relative U(1) charge and the annihilation conditions with 
respect to the fermionic zero modes \Gn\ and \Qn\ (the BRST-invariance 
properties). This classification resulted in: 4 different types of \svec s
for chiral Verma modules built on chiral highest weight vectors
$\ket{0,\htop}^{G,Q}$, 20 different types of \svec s for 
generic (standard) Verma modules (10 types built on 
\Gn-closed h.w. vectors $\ket{\D,\htop}^G$ and 10 types built on
\Qn-closed h.w. vectors $\ket{\D,\htop}^Q$) and 9 different types
of \svec s for no-label Verma modules built on no-label indecomposable
h.w. vectors $\ket{0,\htop}$. In generic Verma modules one can find
\Gn-closed, \Qn-closed, chiral and no-label \svec s. 
In chiral and no-label Verma modules, however, only \Gn-closed and 
\Qn-closed \svec s can exist, with the exception of the chiral \svec s
at level zero in no-label Verma modules (curiously, in chiral and 
no-label Verma modules neither chiral nor no-label \svec s exist).
For the case of the generic Verma modules built on \Gn-closed h.w. vectors
$\ket{\D,\htop}^G$,
which are of special importance for the discussion that follows, the
possible types of \svec s one can find are given by the following table:

\BE
\begin{tabular}{r|l l l l}
{\ }& $q=-2$ & $q=-1$ & $q=0$ & $q=1$\\
\hline\\
\Gn-closed & $-$ & $\kc_l^{(-1)G}$ & $\kc_l^{(0)G}$ & $\kc_l^{(1)G}$\\
\Qn-closed & $\kc_l^{(-2)Q}$ & $\kc_l^{(-1)Q}$ & $\kc_l^{(0)Q}$ & $-$ \\
chiral & $-$ & $\kc_l^{(-1)G,Q}$ & 
$\kc_l^{(0)G,Q}$ & $-$ \\
no-label & $-$ & $\kc_l^{(-1)}$ &
$\kc_l^{(0)}$ & $-$\\
\end{tabular}
\label{tabl2}
\EE
                             
In ref. \cite{BJI6} all \svec s (i.e. 4 + 20 + 9)
were written down explicitly at level 1. This classification
was proved to be rigorous later in ref. \cite{DB2}, where also the 
maximal dimensions of the corresponding singular vector spaces were
given (1, 2 or 3 depending on the type of singular vector). 
Regarding subsingular vectors, in ref. \cite{BJI6} 
all the subsingular vectors in generic Verma modules  
that become singular in the chiral Verma modules were written down 
at levels 2 and 3. To understand this
one has to take into account that chiral Verma modules are nothing but
the quotient of generic Verma modules with zero conformal weight, $\D=0$,
by the submodules generated by the level-zero \svec s (which are present 
in all generic Verma modules with $\D=0$). 

Three months after ref. \cite{BJI6} was published in hep-th/9701041
(January 97), 
the paper `The Structure of Verma Modules over the N=2 Superconformal
Algebra', by Semikhatov and Tipunin, appeared in hep-th/9704111 \cite{SeTi3}.
In this paper the authors considered also the Topological N=2 algebra, 
that they called `the only' N=2 algebra\footnote{Although the 
Topological and the Neveu-Schwarz N=2 algebras are related by the 
topological twists, and the Neveu-Schwarz and the Ramond N=2 algebras are
related by the spectral flows, their representation theories are 
different. This has not been appreciated by the authors who claim that
the corresponding Verma modules are isomorphic, creating some confusion.
In addition, the authors also claim that the Topological and Neveu-Schwarz
N=2 algebras are related by the spectral flows, creating more confusion.
Furthermore there is also the `twisted' N=2 algebra (not to be confused
with the `twisted topological', i.e. the Topological algebra)
which is not connected to the other three algebras.}.
All the analysis and the results presented 
in this paper were based on the following assumptions (without proofs):
 
i) In the N=2 Verma modules there are only two types of submodules. In
particular, in the generic Verma modules built on \Gn-closed h.w. vectors
(called `massive' Verma modules) one can find the two types, denoted 
as `massive' (large) and `topological' (small) submodules.

ii) These two types of submodules are maximally generated (i.e. without 
letting any null states outside, like subsingular vectors) by some 
`non-conventional \svec s', constructed by the authors in 
refs. \cite{SeTi1} \cite{SeTi2}, which satisfy `twisted' h.w. conditions and 
coincide with the conventional \svec s only in the case of `zero twist'.
In more intuitive terms one can think of the `non-conventional' \svec s
simply as certain null states which, unlike the conventional \svec s,
are not located at the bottom of the submodules, that is, they
are not the null states with lowest conformal weight, 
except for the case of `zero twist'.
 
Let us notice already that, although ref. \cite{BJI6} 
appeared in the bibliography given by the authors, the classification of
Verma modules (generic, no-label and chiral),
with their possible existing types of \svec s, was
overlooked. Most surprising, from the very beginning they ignored 
no-label \svec s in the generic Verma modules that they consider
(called `massive' as we pointed out before), as shown in table (1.3), 
which clearly do not fit into the two types of submodules 
described by the authors. 
   
Based on these assumptions the authors presented a `complete' classification
of subsingular vectors (without giving explicit examples) where,
surprisingly, the subsingular vectors given in ref. \cite{BJI6}, which were
the only explicit examples written down so far for the Topological
algebra, did not fit. 
In what follows, in subsections 2.1 and 2.2, we will show that:

i) In generic (`massive') Verma modules one can find {\it at least} four
different types of submodules. Two of them fit, in principle,
into the description of `massive' and `topological' submodules
given by Semikhatov and Tipunin. The other two types do not fit, clearly, 
into that description. 

ii) The subsingular vectors written down in ref. \cite{BJI6} do not
fit into the classification presented by the authors in ref. \cite{SeTi3}, 
providing in fact a proof that the `non-conventional \svec s'  do not
generate maximal submodules since one can find subsingular vectors
outside which are pulled inside the submodule by the action of the
positive lowering operators.

Afterwards, in subsection 2.3, we will argue that the `non-conventional
\svec s', constructed by Semikhatov and Tipunin in the papers \cite{SeTi1}
and \cite{SeTi2}, are poorly defined objects.  They are supposed to be
related to some special types of singular vectors of the affine $\hat sl(2)$
algebra via an isomorphism \cite{FST}\cite{Seiso}. We will also make some 
remarks about this isomorphism and conclude that it is far from satisfactory.
The results discussed in this note
affect profoundly the results presented by Semikhatov and
collaborators in several other publications. This applies very specially
to the classification of N=2 embedding diagrams proposed (although 
never properly done) by Semikhatov and Sirota in the paper \cite{SeSi},
as we point out in subsection 2.4.
In section 3 we will make some final remarks.

\section{The Facts}

\subsection{Different types of submodules}\lvm

In the recent paper \cite{DB4} the determinant formulae for the Topological
N=2 algebra were presented as well as a detailed analysis of the \svec s 
corresponding to the roots of the determinants. In addition it was proved 
-- both theoretically and with explicit examples -- that in generic Verma
modules one can find four different types of submodules just by taking into
account the size and the shape at the bottom of the submodules. Now we will
see that two of these types do not fit into the `massive' and `topological'
submodules claimed to be the only existing
types of submodules by Semikhatov and Tipunin.
The argument goes as follows. The determinant formula for all 
the generic Verma modules  -- either with two h.w. vectors  
$\ket{\D,\htop}^G$ and $\ket{\D,\htop-1}^Q$ ($\D \neq 0$) or with 
only one h.w. vector $\ket{0,\htop}^G$ or $\ket{0,\htop-1}^Q$ --
reads\footnote{The Verma modules built on \Gn-closed h.w. vectors 
and the ones built on \Qn-closed h.w. vectors are not the same
for zero conformal weight $\D=0$ because in this case there is only one
h.w. vector at the bottom of the Verma module together with one \svec . In
some sense there is only one sector in the Verma modules with $\D=0$, as
happens for the Verma modules of the Ramond N=2 algebra for $\D=\ctop/24$.}
 
\BE
det(\cM_l^T)= 
\prod_{2\leq rs \leq 2l}(f_{r,s})^{2 P(l-{rs\over2})}
\prod_{0 \leq k\leq l}(g_k^+)^{2 P_k(l-k)} 
\prod_{0 \leq k\leq l}(g_k^-)^{2 P_k(l-k)}  {\ } ,
\label{det1}
\EE
\noi
where  
\BE
f_{r,s}(\D,\htop,t) = -2t\D + t\htop - \htop^2 - {1 \over 4} t^2
       + {1 \over 4} (s-tr)^2  \,, {\ \ \ }r\in\oZ^+,{\ }s\in2\oZ^+
       \label{frs} \EE
\noi
and 
\BE
g_k^{\pm}(\D,\htop,t) = 2 \D \mp 2k \htop - tk(k \mp 1) \,, {\ \ \ }
	 0 \leq k \in \oZ \,, \label{gk} \EE
\noi
defining the parameter $t={(3-\ctop) / 3}$.
For $\ctop \neq 3 {\ \ } (t \neq 0) {\ }$ one can factorize $f_{r,s}$ as
\BE
f_{r,s}(\D,\htop,t\neq 0) = -2t (\D-\D_{r,s}) \,, \qquad
\D_{r,s} = - {1\over 2t} (\htop-\htop_{r,s}^{(0)}) (\htop-\hat\htop_{r,s})\,,
\label{Drs} \EE
\noi
with
\BE \htop_{r,s}^{(0)} = {t \over 2}(1+r)-{s \over 2} \ , \ \ \ \ \ 
r \in {\bf Z}^+ , \ \ s \in 2{\bf Z}^+ \,, \label{h0rs} \EE
\BE \hat\htop_{r,s} = {t \over 2}(1-r)+{s \over 2}  \ , \ \ \ \ \ 
r \in {\bf Z}^+ , \ \ s \in 2{\bf Z}^+ \,. \label{hhrs} \EE
\noi
For all values of $\ctop$ one can factorize $g_k^+$ and $g_k^-$ as
\BE g_k^{\pm}(\D,\htop,t) = 2 (\D-\D_k^{\pm}) \,, \qquad
\D_k^{\pm} = \pm k \, (\htop-\htop_{k}^{\pm}) \,,  \label{Dk} \EE
\noi
with 
\BE \htop_k^{\pm} =  {t \over 2}\,(1 \mp k ) \,, \ \ \ \ \ k\in {\bf Z}^+ 
\label{hpml} \EE

\noi
The partition functions are defined by
\BE
\sum_N P_k(N)x^N={1\over 1+x^k}{\ }\sum_n P(n)x^n=
{1\over 1+x^k}{\ }\prod_{0<r\in {\bf Z},{\ }0<m\in {\bf Z}}{(1+x^r)^2
 \over (1-x^m)^2} .
\label{part}\EE
The fact that $2 P(0) = 2 P_k(0) = 2$ indicates that the \svec s 
come two by two at the same level, in the same Verma module. 
Generically one is in the $\cG$-sector,
annihilated by (at least) \Gn , while the other is in the $\cQ$-sector, 
annihilated by (at least) \Qn . 
The roots of the quadratic vanishing surface $f_{r,s}(\D,\htop,t)=0$
and of the vanishing planes $g_k^{\pm}(\D,\htop,t)=0$
are related to the corresponding roots of the determinant
formula for the Neveu-Schwarz N=2 algebra 
\cite{BFK}\cite{Nam}\cite{KaMa3}\cite{Yu} via the topological twists. 
These transform the standard h.w. vectors of the 
Neveu-Schwarz N=2 algebra into \Gn-closed h.w. vectors of the
Topological N=2 algebra. As a consequence, under the topological
twists, the Neveu-Schwarz \svec s are transformed into the \svec s 
of the $\cG$-sector of the Topological algebra (see refs.
\cite{BJI6}\cite{BJI7} for a detailed account of the twisting and 
untwisting of primary states and singular vectors).

It is easy to check, by counting of states, that the partitions
$2 P(l-{rs\over2})$, exponents of $f_{r,s}$ in the determinant formula, 
correspond to complete Verma submodules of generic type, whereas
the partitions 
$2 P_k(l-k)$, exponents of $g_k^{\pm}$ in the determinant formula,
correspond to incomplete Verma submodules\footnote{The exponents of the
determinant formulae for the Neveu-Schwarz and for the Ramond N=2 
algebras also show the same behaviour: the exponents of the quadratic
surfaces $f^{A}_{r,s}$ and $f^P_{r,s}$ correspond to complete Verma
submodules and the exponents of the planes $g^A_k$ and $g^P_k$ correspond 
to incomplete Verma submodules, where we use the notation of ref. 
\cite{BFK}. Although this fact is very well known and is even explained 
explicitly in ref. \cite{BFK}, abundantly cited in ref. \cite{SeTi3}, 
Semikhatov and Tipunin believed that they themselves had 
discovered that for N=2 algebras some submodules are `large', of the same
size of the Verma module itself, and some submodules are `small'. See the 
note at the end of ref. \cite{SeTi3}  where they ask for credit for this fact
referring to a paper by M. D\"orrzapf.}. 
Furthermore, as pointed out before, taking into account also 
the shape at the bottom one can distinguish four types of submodules. 
Two of these types correspond to the quadratic vanishing surfaces
$f_{r,s}(\D,\htop,t)=0$, a third type corresponds to the vanishing planes 
$g_{k}^{\pm}(\D, \htop, t)=0$, and the fourth type corresponds to the
`no-label' submodules that one finds in certain intersections of 
$f_{r,s}(\D,\htop,t)=0$ and $g_{k}^{\pm}(\D, \htop, t)=0$, as we will 
explain.

The two types that correspond to the quadratic vanishing surfaces
$f_{r,s}(\D,\htop,t)=0$, have therefore the same size (complete generic 
Verma submodules). The difference between them consists of the shape at 
the bottom, where they both have (in the most general case $\D \neq 0$)
two {\it uncharged} \svec s at level $l={rs\over2}$: $\kc_l^{(0)G}$ in the
$\cG$-sector and $\kc_l^{(0)Q}$ in the $\cQ$-sector\footnote{For the case
$\D=0$ there is only one h.w. vector in the Verma module and therefore
only one of the \svec s can be described as `uncharged' while the other
must necessarily be described as charged with respect to the unique h.w. 
vector. These details are however irrelevant for the present discussion.}.
An important remark now is that if one chooses as h.w. vector of the
Verma module only the \Gn-closed one $\ket{\D,\htop}^G$, as Semikhatov 
and Tipunin do, regarding the \Qn-closed h.w. vector $\ket{\D,\htop-1}^Q$
simply as a descendant state, then the \svec\ $\kc_l^{(0)Q}$ in the 
$\cQ$-sector is necessarily described as a negatively charged state 
$\kc_l^{(-1)Q}$ built on the h.w. vector $\ket{\D,\htop}^G$.
As shown in Figure I, in the most general case the bottom of the submodule 
consists of two \svec s connected by one or two horizontal arrows 
corresponding to \Qn\ and/or \Gn . There is only one arrow if one of the 
\svec s is chiral, i.e. of type $\kc_l^{(0)G,Q}$ instead, what happens
generically for $\D = -l$. These submodules fit, in principle,
into the description of `massive' submodules given by Semikhatov and 
Tipunin. Namely, `massive' submodules are supposed to correspond to 
the uncharged roots $f_{r,s}(\D,\htop,t)=0$, they have the same size as 
the generic (`massive') Verma module and they have two states
at the bottom connected through \Qn\ and/or \Gn , one of these states 
being the \Gn-closed uncharged \svec\ $\kc_l^{(0)G}$ (they do not
mention the possibility that this \svec\ may be chiral for $\D = -l$, though).

It also happens, however, for $\D = -l, \, t = - {s \over n}, \, n=1,..,r$,
that the two \svec s at the bottom of the submodule
are chiral both, and therefore disconnected from each other, as shown
in Fig. I. Consequently these `chiral-chiral' submodules, 
of the same size as the `massive' Verma modules and corresponding
also to the uncharged roots $f_{r,s}(\D,\htop,t)=0$, contain two 
disconnected pieces at the bottom and as a result 
do not fit into the description of `massive' submodules. Nor do they fit 
into the description of two `topological' (smaller) submodules since these
correspond to the charged roots $g_{k}^{\pm}(\D, \htop, t)=0$ with
charged \svec s $\kc_l^{(1)G}$ or $\kc_l^{(-1)G}$ at the bottom of the
submodules, as we will see.

Let us stress that the existence of `chiral-chiral' submodules was obvious 
since January 1997 when the whole set of \svec s of the Topological algebra
at level 1 was written down in ref. \cite{BJI6}. For example, the chiral 
\svec s $\kc_1^{(q)G,Q}$ at level 1 built on \Gn-closed h.w. vectors
$\ket{\D,\htop}^G$ (which are the only h.w. vectors considered by 
Semikhatov and Tipunin) were shown to be:
\BE
\kc_1^{(0)G,Q}=(-2\cL_{-1} + \cG_{-1}\cQ_0 )\ket{-1,-1}^G ,\EE
\BE
\kc_1^{(-1)G,Q}=(\cL_{-1}\cQ_0+\cH_{-1}\cQ_0 +
\cQ_{-1}) \ket{-1,{6-\ctop\over3}}^G . \EE
For $\ctop=9$ ($t=-2$) these two chiral \svec s are together in the same 
generic (`massive') Verma module built on the h.w. vector $\ket{-1,-1}^G$.
Hence these results already prove the existence of
`chiral-chiral' submodules at level 1.

\vskip .3in
\baselineskip 12pt

\vskip .5in
\def\sk {{\hskip 1 cm}}
\def\bk {{\hskip 0.2 cm}}
\def\bbk{{\hskip 0.2 cm}}
\def\rk {{\hskip 3 cm}}
\def\com{{\hskip 0.2 cm},}
\def\pkt{{\hskip 0.2 cm}.}

\def\cQ{{\cal Q}}
\def\cG{{\cal G}}

\def\abs{\,|\,}                                 
\def\tilt {\tilde{t}}             
\def\tils {\tilde{s}}             

\newcommand{\fig}[1]{{\sc Fig.}\,{\sf #1}}        
\newcommand{\figs}[1]{{\sc Figs.}\,{\sf #1}}      
\newcommand{\ch}[1]{\makebox(0,0)[b]{\scriptsize$#1$}}  
\newcommand{\cl}{tt}              
\newcommand{\case}[1]{\makebox(0,0)[b]{\footnotesize #1}}                

\newcounter{pics}
\newcommand{\bpic}[4]{\begin{center}\begin{picture}(#1,#2)(#3,#4)
\refstepcounter{pics}}
\renewcommand{\thepics}{{\sf\Roman{pics}}}
\newcommand{\epic}[1]{\end{picture}\\
\end{center}{ \footnotesize {\sc Fig.} \thepics \bk #1}}
\newcommand{\epicspl}{\end{picture}\\           
\addtocounter{pics}{-1}\end{center}}
\baselineskip 12pt              

\vbox{
\bpic{400}{120}{20}{-20}

\put(48,20){\framebox(4,3){}}
\put(78,20){\framebox(4,3){}}
\put(54,23){\vector(1,0){22}}
\put(76,20){\vector(-1,0){22}}
\put(47,24){\line(-1,2){50}}
\put(82,24){\line(1,2){50}}


\put(51,51){\circle*{3}}
\put(79,51){\circle*{3}}
\put(54,49){\vector(1,0){22}}
\put(76,52){\vector(-1,0){22}}
\put(49,53){\line(-1,2){30}}
\put(81,53){\line(1,2){30}}

\put(65,-10){\case{$f_{r,s}(\D,\htop,t)=0$, $\Delta \neq - l$}}


\put(198,20){\framebox(4,3){}}
\put(228,20){\framebox(4,3){}}
\put(204,23){\vector(1,0){22}}
\put(226,20){\vector(-1,0){22}}
\put(197,24){\line(-1,2){50}}
\put(232,24){\line(1,2){50}}
\put(201,51){\circle*{3}}
\put(229,51){\circle*{3}}
\put(226,51){\vector(-1,0){22}}
\put(199,53){\line(-1,2){30}}
\put(231,53){\line(1,2){30}}
\put(215,-10){\case{$f_{r,s}(\D, \htop, t)=0$, $\Delta = - l$, 
$t \neq - {s \over n}$}}

\put(348,20){\framebox(4,3){}}
\put(378,20){\framebox(4,3){}}
\put(354,23){\vector(1,0){22}}
\put(376,20){\vector(-1,0){22}}
\put(347,24){\line(-1,2){50}}
\put(382,24){\line(1,2){50}}
\put(351,51){\circle*{3}}
\put(379,51){\circle*{3}}
\put(349,53){\line(-1,2){30}}
\put(381,53){\line(1,2){30}}
\put(353,53){\line(1,2){30}}
\put(377,53){\line(-1,2){30}}
\put(365,-10){\case{$f_{r,s}(\D, \htop, t)=0$, $\Delta = - l$, 
$t = - {s \over n}$ }}
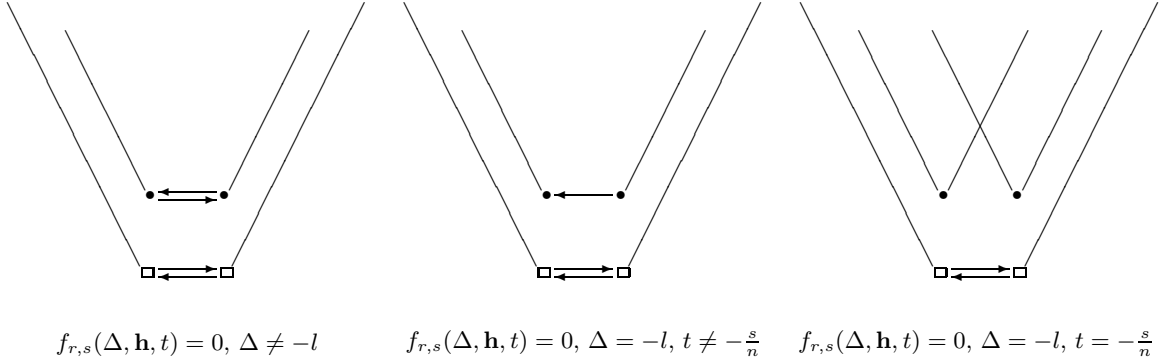
\epic{. The \svec s corresponding to the series $f_{r,s}(\D, \htop, t)=0$
belong to two different types of submodules of the same size (complete
Verma submodules). In the first type, as shown in the figures on the
left and in the center, the two \svec s at the bottom of
the submodules are connected by \Gn\ and/or \Qn , depending on whether
$\D \neq -l$ or $\D = -l, \, t \neq - {s \over n}, \, n=1,..,r $ (for which 
one of the \svec s is chiral). In the second type, corresponding to 
$\D = -l, \, t = - {s \over n}, \, n=1,..,r $, the two \svec s 
are chiral and therefore disconnected from each other, as shown in the
figure on the right.}}

\baselineskip 16pt
\vskip .2in

The third type of submodules, shown in Fig. II, corresponds to the roots 
of the vanishing 
planes $g_{k}^{\pm}(\D, \htop, t)=0$. As already pointed out, these are
smaller, incomplete Verma submodules with partition functions $P_k(l-k)$. 
In the most general case ($\D \neq 0$) the
two \svec s at the bottom of the submodule can be described as 
{\it charged}: positively charged $\kc_l^{(1)G}$ in the $\cG$-sector and  
$\kc_l^{(1)Q}$ in the $\cQ$-sector for $g_{k}^{+}(\D, \htop, t)=0$, and
negatively charged $\kc_l^{(-1)G}$ in the $\cG$-sector and  
$\kc_l^{(-1)Q}$ in the $\cQ$-sector for $g_{k}^{-}(\D, \htop, t)=0$. In 
each case one of the \svec s becomes chiral for $\D=-l$ whereas the other
\svec\ does not. The bottom of these submodules is always connected 
therefore as there is no analog to the `chiral-chiral' case of the 
uncharged \svec s. When the bottom of the submodule is at level zero in
the Verma module, then there is only one \svec , which consequently is chiral. 
These submodules seem to fit well the description of `topological' 
submodules given by Semikhatov and Tipunin. Namely, they correspond to 
the charged roots $g_{k}^{\pm}(\D, \htop, t)=0$, they are smaller than the 
`massive' submodules and they have one or two states at the bottom. In
the first case the unique state is charged and chiral (called `topological') while
in the second case the two states are connected through \Gn\ and/or \Qn , 
one of these states being the \Gn-closed charged \svec\ $\kc_l^{(1)G}$ (for 
$g_{k}^+(\D, \htop, t)=0$) or $\kc_l^{(-1)G}$ (for $g_{k}^-(\D, \htop, t)=0$).

Finally, the fourth type of submodules, shown also in Fig. II, corresponds 
to the `no-label' submodules. These are the widest submodules, with four
\svec s at the bottom, generated
by `no-label' \svec s. These are {\it primitive}
\svec s that only exist for discrete values of
$\D, \htop, t$, in Verma modules where there are intersections,
at the same level $l$, of \svec s corresponding to the series 
$f_{r,s}(\D, \htop, t)=0$ with \svec s corresponding to one of the series
$g_k^{\pm}(\D, \htop, t)=0$, with ${rs \over 2} = k = l$ and $\D = -l$.
The values of $\ctop$ for which no-label \svec s exist are
$\ctop = {3r-6 \over r}\,$, corresponding to $t={2 \over r}$. 
These results have been proved in ref. \cite{DB4} although
the existence of no-label \svec s was proved 
in January 1997 in ref. \cite{BJI6}, since
they were explicitly written down at level 1 (shortly afterwards no-label
\svec s were written down at level 2 in ref. \cite{DB1} -- this was even
advertized in the abstract of the paper-- where it was proved that no-label
\svec s of the Topological algebra correspond to subsingular vectors of
the Neveu-Schwarz N=2 algebra under the topological twists). 

\vskip .5in

\baselineskip 12pt
\vbox{
\bpic{400}{120}{20}{-20}


\put(48,20){\framebox(4,3){}}
\put(78,20){\framebox(4,3){}}
\put(54,23){\vector(1,0){22}}
\put(76,20){\vector(-1,0){22}}
\put(47,24){\line(-1,2){50}}
\put(82,24){\line(1,2){50}}

\put(81,81){\circle*{3}}
\put(99,81){\circle*{3}}
\put(84,79){\vector(1,0){12}}
\put(96,82){\vector(-1,0){12}}
\put(79,83){\line(-1,2){20}}
\put(101,83){\line(1,2){20}}

\put(65,-10){\case{$g^{\pm}_k(\Delta,\htop,t)=0$, $\D \neq -l$}}
\put(198,20){\framebox(4,3){}}
\put(228,20){\framebox(4,3){}}
\put(204,23){\vector(1,0){22}}
\put(226,20){\vector(-1,0){22}}
\put(197,24){\line(-1,2){50}}
\put(232,24){\line(1,2){50}}

\put(231,81){\circle*{3}}
\put(249,81){\circle*{3}}
\put(234,79){\vector(1,0){12}}
\put(229,83){\line(-1,2){20}}
\put(251,83){\line(1,2){20}}

\put(215,-10){\case{$g^{\pm}_k(\Delta,\htop,t)=0$, $\D = -l$}}

\put(348,20){\framebox(4,3){}}
\put(378,20){\framebox(4,3){}}
\put(354,23){\vector(1,0){22}}
\put(376,20){\vector(-1,0){22}}
\put(347,24){\line(-1,2){50}}
\put(382,24){\line(1,2){50}}

\put(361,80){\circle*{3}}
\put(380,78){\circle*{3}}
\put(380,82){\circle*{3}}
\put(399,80){\circle*{3}}
\put(384,78){\vector(1,0){12}}
\put(396,82){\vector(-1,0){12}}
\put(364,82){\vector(1,0){12}}
\put(376,78){\vector(-1,0){12}}
\put(359,83){\line(-1,2){20}}
\put(401,83){\line(1,2){20}}

\put(365,-10){\case{no-label submodules}}
\epic{. 
The \svec s
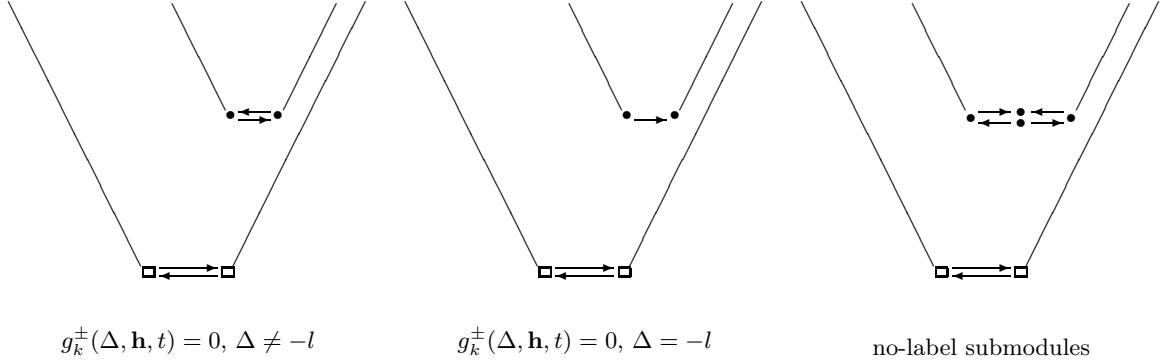
corresponding to the series $g_{k}^{\pm}(\D, \htop, t)=0$ belong to only
one type of submodules (incomplete Verma submodules). In the 
generic case $k \neq 0$ there are two \svec s at the bottom of the 
submodules, connected by \Gn\ and/or \Qn , depending on whether 
$\D \neq -l$ or $\D=-l$. (For $k=0$, that is level zero, the bottom
of the submodule consists of only one \svec\ which is chiral). 
The no-label \svec s generate the widest submodules with four
\svec s at the bottom. }}

\vskip .2in

\baselineskip 16pt

The action of \Gn\ and \Qn\ on a no-label \svec\ $\kc_l^{(q)}$ produce three 
secondary singular vectors (one \Gn-closed, one \Qn-closed and one chiral)
which cannot `come back' to the no-label \svec\ 
by acting with \Gn\ and \Qn :
 \begin{eqnarray} \Qz \,\kcc{l}{q}{ } \to \kcc{l}{q-1}{Q} ,  
 \quad\Gz \,\kcc{l}{q}{ } \to \kcc{l}{q+1}{G} , \quad 
 \Gz \, \Qz \,\kcc{l}{q}{ } \to \kcc{l}{q}{G,Q} \,. \label {QGnh} 
 \end{eqnarray}
\noi
It happens that one of these \svec s 
corresponds to the series $f_{r,s}(\D, \htop, t)=0$, another one 
corresponds to the series $g_k^{\pm}(\D, \htop, t)=0$, and the remaining 
one corresponds to both series. 
Hence the bottom of the no-label submodules is connected,
generated by the no-label \svec\ and consists of four \svec s: the primitive
no-label \svec\ and the three secondary \svec s. Obviously, these submodules
are wider than the `massive' submodules (twice wider at the bottom, 
in fact) and do not fit into the description of `massive' and 
`topological' submodules.

In  Fig. III one can see the case of an uncharged no-label \svec\ 
$\ket{\chi}_{l}^{(0)}$ with the three corresponding secondary \svec s.
The uncharged no-label \svec\ $\kc_1^{(0)}$ at level 1,
built on a \Gn-closed h.w. vector $\ket{\D, \htop}^G$,
together with the three secondary \svec s that it generates
at level 1 by the action of \Gn\ and \Qn\ read:
\BE \kc_{1,\ket{-1,-1, \, t=2}^G}^{(0)}\, = 
(\cL_{-1} - \cH_{-1})\ket{-1,-1, \, t=2}^G ,\EE
\BE \kc_{1,\ket{-1,-1, \, t=2}^G}^{(1)G}=
\cG_0\kc_{1,\ket{-1,-1, \, t=2}^G}^{(0)}\, =
 2 \cG_{-1} \ket{-1,-1, \, t=2}^G ,\EE
\BE \kc_{1,\ket{-1,-1, \, t=2}^G}^{(-1)Q}=
\cQ_0\kc_{1,\ket{-1,-1, \, t=2}^G}^{(0)}\,
 = (\cL_{-1}\cQ_0 - \cH_{-1}\cQ_0 - \cQ_{-1})\ket{-1,-1, \, t=2}^G ,\EE 
\BE  
\kc_{1,\ket{-1,-1, \, t=2}^G}^{(0)G,Q} = 
 \cG_0\cQ_0\kc_{1,\ket{-1,-1, \, t=2}^G}^{(0)} = 
 2 (-2\cL_{-1} + \cG_{-1}\cQ_0 )\ket{-1,-1, \, t=2}^G .\EE
\noi
The no-label \svec\ only
exists for $t=2$ ($\ctop=-3$) whereas the three secondary \svec s are just
the particular cases, for $t=2$, of the one-parameter families of \svec s
of the corresponding types, which exist for all values of $t$ \cite{BJI6}. 

\vskip .5in
\baselineskip 12pt
\vbox{
\bpic{100}{120}{20}{-20}
\put(48,20){\framebox(4,3){}}
\put(78,20){\framebox(4,3){}}
\put(54,24){\vector(1,0){22}}
\put(76,19){\vector(-1,0){22}}
\put(47,24){\line(-1,2){50}}
\put(82,24){\line(1,2){50}}
\put(65,14){\makebox(0,0){\scriptsize $\cQ_0$}}
\put(65,28){\makebox(0,0){\scriptsize $\cG_0$}}
\put(47,15){\makebox(0,0)[r]{\scriptsize $\ket{-l,\htop-1}^Q$}}
\put(83,15){\makebox(0,0)[l]{\scriptsize $\ket{-l,\htop}^G$}}

\multiput(80,24) (0,5) {14}{\circle*{.5}}
\multiput(78,24) (-2,5) {14}{\circle*{.5}}
\multiput(81,24) (2,5) {14}{\circle*{.5}}

\put(51,91){\circle*{3}}
\put(80,89){\circle*{3}}
\put(80,93){\circle*{3}}
\put(108,91){\circle*{3}}
\put(76,89){\vector(-1,0){22}}
\put(54,93){\vector(1,0){22}}
\put(84,89){\vector(1,0){22}}
\put(106,93){\vector(-1,0){22}}
\put(101,76){\makebox(0,0)[rb]{\scriptsize $\ket{\chi}_{l}^{(0)}$}}
\put(50,89){\makebox(0,0)[rb]{\scriptsize 
	$\ket{\chi}_{l}^{(-1)Q}$}}
\put(116,82){\makebox(0,0)[lb]{\scriptsize 
	$\ket{\chi}_{l}^{(1)G}$}}
\put(60,94){\makebox(0,0)[lb]{\scriptsize 
	$\ket{\chi}_{l}^{(0)G,Q}$}}
\epic{. 
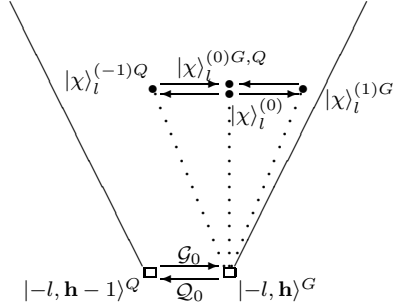
The uncharged no-label \svec\ $\ket{\chi}_{l}^{(0)}$ at level $l$, built
on the h.w. vector $\ket{-l,\htop}^G$, is the primitive \svec\ generating
the three secondary \svec s at level $l$:
$\ket{\chi}_{l}^{(1)G} = \cG_0 \ket{\chi}_{l}^{(0)}$,
$\ket{\chi}_{l}^{(-1)Q} = \cQ_0 \ket{\chi}_{l}^{(0)}$ and  
$\ket{\chi}_{l}^{(0)G,Q} = \cQ_0 \cG_0 \ket{\chi}_{l}^{(0)}
= - \cG_0 \cQ_0 \ket{\chi}_{l}^{(0)}$. These cannot generate the no-label
\svec\ by acting with the algebra. However, they are the \svec s detected
by the determinant formula, corresponding to the series 
$f_{r,s}(\D, \htop, t)=0$ ($\ket{\chi}_{l}^{(-1)Q}$ and
$\ket{\chi}_{l}^{(0)G,Q}$) and the series $g_{k}^+(\D, \htop, t)=0$
($\ket{\chi}_{l}^{(1)G}$ and $\ket{\chi}_{l}^{(0)G,Q}$). }}

\baselineskip 16pt
\vskip .2in

Two important observations come now in order. First, a given submodule 
may not be completely generated by the \svec s at the bottom. 
These could generate only a submodule of the whole (maximal) submodule,
in which case 
one or more {\it subsingular} vectors generate the missing parts. 
Second, we have seen that there are four different types of submodules
that may appear in generic Verma modules, distinguished by their size
and/or the shape at the bottom of the submodule.
A more accurate classification of the submodules, however, 
should take into account also the shape of the whole  
submodule, including the possible existence of subsingular 
vectors \cite{DB5}. For this reason we claim that there are {\it at least}
four different types of submodules in generic Verma modules.

We have shown so far that the two types of submodules proposed by 
Semikhatov and Tipunin -- `massive' and `topological' submodules -- 
correspond, in principle, to the submodules of the first and third types, 
respectively, whereas the submodules of the second and fourth types 
(`chiral-chiral' and `no-label' submodules) have been overlooked by 
the authors. (As a matter of fact, it is not clear if chiral {\it uncharged} 
\svec s $\ket{\chi}_{l}^{(0)G,Q}$ fit in the framework of the authors 
at all since they are completely ignored). 
Furthermore the authors claim that the `massive' and 
`topological' submodules are generated {\it maximally} (i.e. 
completely, without letting any null states outside) by the non-conventional 
\svec s that they constructed in refs. \cite{SeTi1}\cite{SeTi2}. 

Let us notice that the very existence of no-label \svec s already disproves this 
claim because no-label \svec s (and many of their descendants) can be viewed 
as pieces left outside from the `topological' and `massive' submodules.  
To be precise, no-label submodules contain one `topological' and one
`massive' submodule, starting at the bottom, but the no-label \svec ,
and many of its descendants, do not belong to these submodules. As a result
the `massive' and `topological' submodules together do not build the
whole no-label submodule. Consequently, the non-conventional \svec s
together do not generate the maximal no-label submodule 
(not even the bottom!). In the next subsection we will give another argument 
showing in a different way that the non-conventional \svec s do not generate
maximal submodules: one can find subsingular vectors outside of them!

\subsection{The classification of subsingular vectors}\lvm

In order to understand the results presented by Semikhatov and Tipunin in 
ref. \cite{SeTi3} (and in several other publications)
we have to make two important remarks concerning the 
presentation of the {\it conventional} \svec s by the authors.
First, the authors claim that in the {\it conventional} approach the h.w. conditions
imposed on the h.w. vectors and on {\it any} \svec\ must include the annihilation
by \Gn\ (eq.(2.11) in ref. \cite{SeTi3}). This statement is incorrect since 
in the conventional approach, for the (super)conformal algebras,
one defines the h.w. vectors and \svec s (often called simply null vectors) 
as the states with lowest conformal weight (lowest energy) in the Verma 
modules and submodules, respectively. As a result, in most Verma modules and 
submodules of the Ramond and the Topological N=2 algebras (they are 
isomorphic in fact \cite{DB4}) there are two sectors degenerated in energy, the 
$+$ and $-$ sectors for the Ramond algebra and the $\cG$ and $\cQ$ sectors 
for the Topological algebra, the corresponding states annihilated by the 
fermionic zero modes $G^+_0$ or $G^-_0$ and \Gn\ or \Qn , respectively
\cite{BFK}\cite{KaMa3}\cite{Yu}\cite{Kir1}\cite{LVW}\cite{MSS}\cite{DB4}.  
That is, at the bottom of most Verma modules and submodules of 
the Ramond and of the Topological N=2 algebras there are two h.w. vectors and 
two \svec s, respectively, the fermionic zero modes interpolating between them.
In addition, one 
can find indecomposable \svec s not annihilated by any of the fermionic zero 
modes, that also must be called \svec s following the conventional definition
\cite{BJI6}\cite{DB1}\cite{DB2}\cite{DB4}.

Second, let us also notice that to break the symmetry between the $\cG$ 
and the $\cQ$ sectors, regarding the \svec s of the $\cQ$-sector simply as 
descendant states (non-singular) of `the \svec s' of the $\cG$-sector, leads 
to confusion in the case of zero conformal weight $\D+l=0$. The reason 
is that for $\D+l=0$ the \Qn-closed (non-chiral) \svec s
$\ket{\chi}_{l=-\D}^{(q)Q}$ are in fact the primitive ones generating the 
{\it secondary} \svec s of the $\cG$-sector, which are necessarily chiral 
of type $\ket{\chi}_{l=-\D}^{(q+1)G,Q}$ (see the details in ref. \cite{DB4},
Appendix A). In the conventions used by Semikhatov and Tipunin, however,
the vectors $\ket{\chi}_{l=-\D}^{(q)Q}$ are not singular by definition. As a
result, since they are not descendant states of `the \svec ' 
$\ket{\chi}_{l=-\D}^{(q+1)G,Q}$, but the other way around, 
the \svec s of the $\cQ$-sector $\ket{\chi}_{l=-\D}^{(q)Q}$ 
must be called {\it subsingular vectors} instead. For similar reasons,
the indecomposable `no-label' \svec s must also be called subsingular
vectors (they are not descendants of the \svec s of the $\cG$-sector, but
the other way around, and they are not singular by definition). 

Now we will see that the explicit examples of subsingular vectors given
in ref. \cite{BJI6}, which are singular in the chiral Verma modules, 
do not fit into the complete classification of subsingular vectors 
presented by Semikhatov and Tipunin
in ref. \cite{SeTi3}. As a consequence we will deduce that the 
non-conventional \svec s constructed in refs.
\cite{SeTi1}\cite{SeTi2} do not generate maximal submodules.
The authors classified the generic Verma modules built on \Gn-closed
h.w. vectors (`massive' Verma modules) 
according whether they have zero, one, two or more \svec s from the 
uncharged and/or charged series associated to the roots of the determinant 
formula (in our notation $f_{r,s}(\D, \htop, t)=0$ and/or 
$g_k^{\pm}(\D, \htop, t)=0$). In every case they applied the assumption 
that there are only two types of submodules -- `massive' and `topological' 
-- and these are generated maximally by the non-conventional \svec s
constructed in refs. \cite{SeTi1}\cite{SeTi2}. 
Namely, one `twisted topological' non-conventional \svec\
(where they mean twisted by the spectral flows) 
is assumed to generate maximally one `topological' submodule 
whereas one `twisted massive' non-conventional \svec\ 
is assumed to generate maximally one `massive' submodule. 

As pointed out before, these objects are 
null states that in general are not located at the bottom of the 
submodules unlike the conventional \svec s. In fact, in the cases when
they lie at the bottom then they coincide with the conventional \svec s. An 
important remark is that the `twisted topological' h.w. conditions satisfied
by the `twisted topological' non-conventional \svec s reduce to the
chirality h.w. conditions (i.e. annihilation by \Gn , \Qn\ and by all the
positive generators) in the case of the twist parameter equal to zero.
As a result, the 'zero twist topological' non-conventional \svec s
coincide with the chiral {\it charged} conventional \svec s at the
bottom of the `topological' submodules. 
The claim that the non-conventional \svec s generate
maximal submodules implies that acting with the lowering and 
raising operators of the algebra one 
generates whole submodules without any null states left outside, 
such as subsingular vectors and descendants of them, that can be pulled
inside the submodules by the action of the algebra.

Using these assumptions and simple geometrical arguments, the authors
deduced in which cases the {\it conventional} \svec s at the bottom of the
submodules do not generate maximal submodules, the remaining pieces
outside being generated by subsingular vectors. In some of these cases the
authors gave general expressions for the subsingular vectors.   
The subsingular vectors given by us in ref. \cite{BJI6} must correspond
necessarily to the ones described by the authors in the case
`codimension-2 charge-massive', given\footnote{The authors themselves
claimed that the subsingular vectors given in ref. \cite{BJI6} were
described by Proposition 3.9, case $n=0$ \cite{Tip}, although they did not 
explicitely mention this in the last revised, published version of
ref. \cite{SeTi3}. See a comment in the Final Remarks regarding this issue.}
in Proposition 3.9, for $n=0$, since they are located in Verma modules
with one charged chiral \svec\ (at level zero, what gives $n=0$) and one 
uncharged \Gn-closed \svec\ (and its companion in the $\cQ$-sector that
is ignored by the authors). In the notation of the authors, 
who draw the Verma modules upside-down, 
the charged \svec\ is both a conventional `top-level' \svec\ and a 
non-conventional `twisted topological' charged \svec\ 
$\ket{E(n)}_{ch}$ with twist parameter $n=0$ 
(i.e. the non-conventional \svec\ is at the bottom of
the submodule so that it coincides with the conventional \svec ). 
The uncharged \Gn-closed \svec\ is described as the conventional 
`top-level' uncharged \svec\ in the `massive' submodule generated 
by the `massive' \svec\  $\ket{S(r,s)}$ , and is denoted as $\ket{s}$.

For this case, and in fact for all cases `described' by Proposition 3.9,
the authors deduced that a subsingular vector $\ket{Sub}$ must exist
{\it inside} the maximal massive submodule generated by $\ket{S(r,s)}$
in the sense that $\ket{Sub}$ is located {\it outside} the non-maximal
submodule generated by the conventional uncharged \svec\ $\ket{s}$, 
becoming singular once $\ket{s}$ is set to zero. This implies
that the subsingular vector $\ket{Sub}$ is `pushed down' (`up' in the 
authors figures) by the action of the lowering operators inside the 
non-maximal submodule generated by $\ket{s}$, so that setting
this submodule to zero is equivalent to push down the vector to nothing,
i.e. the subsingular vector becomes singular. Observe that in this case
the subsingular vector $\ket{Sub}$, once it reaches $\ket{s}$ by
the action of the lowering operators, cannot go down (`up') 
anymore since $\ket{s}$ is the conventional \svec\ at the bottom of
the submodule annihilated by all the lowering operators. In other words,
if the subsingular vector $\ket{Sub}$ becomes singular when $\ket{s}$ is 
set to zero, then acting with the lowering operators on $\ket{Sub}$ it
cannot be pulled down beyond the level of $\ket{s}$, getting in fact 
`stuck' in $\ket{s}$ (up to constants).

The subsingular vectors at level 3 given by us in ref. \cite{BJI6} do not
follow the behaviour described by Proposition 3.9, however. Rather, they
are pulled down beyond the uncharged conventional \svec\ $\ket{s}$ 
that one finds at level 2 and, in fact, they can be pulled down until
the very end, i.e. level zero, becoming singular only when the charged 
chiral \svec\ $\ket{E(0)}_{ch}$ at level zero is set to zero. As a
consequence, these subsingular vectors do not become singular when 
$\ket{s}$ is set to zero, what implies that they are not pulled inside the 
submodule generated by $\ket{s}$ by the action of the lowering operators
(see Fig. IV), 
and therefore they are not located inside the maximal massive submodule 
supposed to be generated by the `massive' \svec\ $\ket{S(r,s)}$. But these 
subsingular vectors are neither located inside the submodule generated 
by $\ket{E(0)}_{ch}$ since they do not disappear when $\ket{E(0)}_{ch}$
is set to zero, becoming singular rather. In other words, 
as shown in Fig. IV, these subsingular
vectors are pulled inside the submodule generated by $\ket{E(0)}_{ch}$
by acting with the lowering operators. This implies that the submodule
generated by the non-conventional \svec\ $\ket{E(0)}_{ch}$ is not maximal,
in contradiction with the claims of Semikhatov and Tipunin.   
 
One example given in ref. \cite{BJI6} is the subsingular vector 
$\ket{Sub}_3^{(1)}$ at level 3 with charge $q=1$ built on the \Gn-closed
h.w. vector $\ket{\D, \htop}^G$ with conformal weight $\D=0$ and U(1) 
charge $\htop=2$:
\begin{eqnarray*}
\ket{Sub}_3^{(1)}= \{ {3-\ctop\over 24} {\ } \cL_{-1}^2 \cG_{-1} - {3\over 4} {\ }
\cL_{-1} \cG_{-2} - {1\over 4} {\ } \cL_{-2} \cG_{-1} + 
{\ctop+9\over 4(\ctop-3)} {\ } \cH_{-2} \cG_{-1} +  \\
{27 - \ctop\over 4(3-\ctop)}  {\ } \cG_{-3} + {6\over\ctop-3} {\ } \cH_{-1} \cG_{-2} + 
{3\over 4} {\ } \cH_{-1} \cL_{-1} \cG_{-1} + 
{3\over 3-\ctop} {\ } \cH_{-1}^2 \cG_{-1} \} {\ } \ket{0,2}^G \,.
\end{eqnarray*}
\noi
Acting with $\cQ_1$ on this vector one does not hit the conventional
uncharged \svec\ $\ket{s}$ at level 2 but one reaches the state
\BE  \{ {\ctop-12\over 12} {\ } \cL_{-1} \cG_{-1} + 
{3(11-\ctop)\over 4(3-\ctop)}  {\ } \cG_{-2} +
{3(11-\ctop)\over 4(\ctop-3)}  {\ } \cH_{-1} \cG_{-1} \} {\ } \cQ_0 {\ } \ket{0,2}^G \,,
\EE
\noi
which is a non-singular descendant of the level zero charged \svec\  
$\,\ket{E(0)}_{ch} = \cQ_0 \, \ket{0,2}^G $. That is, $\ket{Sub}_3^{(1)}$ is
pulled inside the submodule generated by $\ket{E(0)}_{ch}$ by the 
action of $\cQ_1$. Acting further with $\cL_1$ 
one reaches the state $\, \cG_{-1} \cQ_0 \, \ket{0,2}^G \,$ at level 1
which, again, is not singular. Acting with $\cQ_1$ on this state one
reaches finally the level zero chiral charged \svec :
$\,\cQ_1\cL_1\cQ_1 \, \ket{Sub}_3^{(1)} = \cQ_0 \, \ket{0,2}^G = \ket{E(0)}_{ch}$.

\vskip .8in
\baselineskip 12pt
\vbox{
\bpic{250}{120}{20}{-20}



\put(100,20){\circle*{.5}}
\put(128,20){\framebox(4,3){}}
\put(126,20){\vector(-1,0){22}}
\put(97,24){\line(-1,2){50}}
\put(132,24){\line(1,2){50}}
\put(115,14){\makebox(0,0){\scriptsize $\cQ_0$}}
\put(97,15){\makebox(0,0)[r]{\scriptsize $\ket{E(0)}_{ch} = \cQ_0\ket{0,2}^{G}$}}
\put(133,15){\makebox(0,0)[l]{\scriptsize $\ket{0,2}^{G}$}}
\put(101,24){\line(1,6){16}}
\put(99,24){\line(-1,3){32}}

\put(103,80){\makebox(0,0)[rb]{\scriptsize $\ket{s}$}}
\multiput(100,75) (1,1) {39}{\circle*{.5}}
\multiput(98,73) (-1,0) {9}{\circle*{.5}}
\multiput(88,75) (-1,2) {18}{\circle*{.5}}
\put(138,113){\circle*{3}}
\put(155,115){\makebox(0,0)[rb]{\scriptsize $\ket{S(r,s)}$}}


\put(129,87){\circle*{3}}
\put(126,85){\vector(-1,-1){29}}
\put(159,89){\makebox(0,0)[rb]{\scriptsize $\ket{Sub}_3^{(1)}$}}
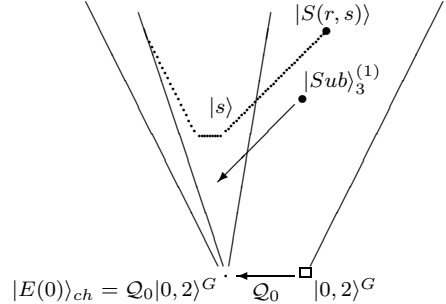
\epic{. When the charged level zero \svec\ $\ket{E(0)}_{ch} = \cQ_0\ket{0,2}^G$
is set to zero, the generic (`massive') Verma module $V(\ket{0,2}^G)$ is divided 
by the submodule generated by this \svec . 
As a result one obtains the incomplete, chiral Verma module 
$V(\ket{0,2}^{G,Q})$ built on the chiral h.w. vector $\ket{0,2}^{G,Q}$. 
The subsingular vector $\ket{Sub}_3^{(1)}$ at level 3 is 
outside the submodule generated by $\ket{E(0)}_{ch}$, being
pulled inside by the action of the lowering operators. 
Consequently, the submodule generated by the non-conventional `topological' 
charged \svec\ $\ket{E(0)}_{ch}$ (which being at the bottom of the
submodule coincides with the conventional chiral \svec\ $\cQ_0\ket{0,2}^G$)
is not maximal since there is (at least) one subsingular vector left outside.
This subsingular vector becomes singular, therefore,
in the chiral Verma module $V(\ket{0,2}^{G,Q})$ obtained after the quotient.
Inside the submodule generated by $\ket{E(0)}_{ch}$ one finds the 
uncharged \Gn-closed \svec\ $\ket{s}$ (and its companion in the 
$\cQ$-sector that is not indicated). The subsingular vector $\ket{Sub}_3^{(1)}$
is not pulled inside the submodule generated by  $\ket{s}$ by
the lowering operators and therefore it does not become singular once 
$\ket{s}$ is set to zero. As a result $\ket{Sub}_3^{(1)}$ does not belong
to the `massive' submodule, supposed to be generated by the non-conventional
`massive'  \svec\ $\ket{S(r,s)}$, having $\ket{s}$ at the bottom. }}

\vskip .18in
\baselineskip 16pt

This example not only proves that Proposition 3.9 is incorrect, as 
$\ket{Sub}_3^{(1)}$ does not become singular when $\ket{s}$ is set to  
zero, and that the subsingular vectors presented in ref. \cite{BJI6}
(the only examples known at that time!) do not fit into the
`complete' classification of subsingular vectors given by Semikhatov 
and Tipunin in ref. \cite{SeTi3}. As we have just discussed, this
example also proves that the non-conventional topological \svec\ 
$\,\ket{E(0)}_{ch} = \cQ_0  \ket{0,2}^G $ (which is located at the
bottom of the submodule and therefore coincides with the conventional 
chiral \svec ) does NOT generate a maximal submodule since the
subsingular vector $\ket{Sub}_3^{(1)}$ is outside this submodule, 
being pulled inside by the action of the lowering operators. 
This example disproves the claim of Semikhatov and Tipunin that
their non-conventional `massive' and `topological' \svec s generate
maximal submodules with no space left outside for subsingular vectors.
Indeed, we have shown that the subsingular vector           
$\ket{Sub}_3^{(1)}$ is neither generated by the `massive' \svec\ 
$\ket{S(r,s)}$ nor by the `topological' \svec\ $\ket{E(0)}_{ch}$, nor
by both of them together. As a result, this example proves that 
the non-conventional `massive' and `topological' \svec s constructed 
by Semikhatov and Tipunin do not generate maximal submodules.

\subsection{The non-conventional \svec s and the isomorphism N=2
 $\leftrightarrow \hat sl(2)$}\lvm

We will argue now that the non-conventional \svec s of Semikhatov and 
Tipunin, apart from the fact that they do not generate maximal submodules, are 
poorly defined objects of unclear meaning (except for some simple cases). 

On the one hand, the key idea underlying the construction of these
`non-conventional' \svec s in refs. \cite{SeTi1} and \cite{SeTi2} was the
misconception that the spectral flows map h.w. vectors to h.w. vectors
(for any value of the parameter $\theta$!) transforming the Verma modules 
into isomorphic Verma modules, consequently\footnote{This is also the
reason why they believed that the Verma modules of the Neveu-Schwarz
and the Ramond N=2 algebras were isomorphic. We tried to warn 
the authors, unsuccesfully, about their wrong use of the
spectral flows. We also suggested to them to use the involutive
automorphism $\cA$ that we had deduced in ref. \cite{BJI3}, eq. (2.8), 
acting as $\cL_m \to \cL_m-m\cH_m$, $\cH_m \to -\cH_m-{\ctop\over3}$,
$\cG_n \to \cQ_n$ and $\cQ_n \to \cG_n$, with $\cA^{-1}=\cA$. This 
suggestion was followed by the authors, but without citing our work, as the 
reader can see in ref. \cite{SeTi2}, eq. (2.7), and in ref. \cite{FSST}, 
eq. (2.17).}.  (The fact that the spectral flows do not map h.w. vectors 
to h.w. vectors, except for some specific values of $\theta$, 
was already pointed out by Schwimmer and Seiberg just a few lines after
they wrote down the spectral flows in ref. \cite{SS}, abundantly cited
by Semikhatov and Tipunin. An exhaustive analysis of this issue for the
even and the odd spectral flows can be found in ref. \cite{B1}). 
To be precise, the authors claimed that
the spectral flows do not act only on the states but also on the h.w.
conditions in such a way that the h.w. vectors always remain h.w. vectors
under any spectral flow transformation. (Observe that transforming the
states and the observables, like the h.w. conditions, at the same time
is equivalent to not doing any transformation at all, just a redefinition
of the states in the same Verma module).   
One can see this misconception abundantly used in ref. \cite{SeTi1}. 
In the last version of ref. \cite{SeTi2} (Sept. 1998), however, the
corresponding misleading statements have been removed everywhere in the
paper. Intriguingly enough, all the final expressions and results have remained
the same.  

On the other hand, the `non-conventional' \svec s are constructed out
of `continued' operators or `intertwiners', 
$g(a,b)$ and $q(a,b)$, that generalize the products of fermionic modes
$\cG_a \cG_{a+1} \cG_{a+2} ...... \cG_{a+N}$ and
$\cQ_a \cQ_{a+1} \cQ_{a+2} ...... \cQ_{a+N}$, respectively, to a {\it complex}
number of factors. The authors claim that this procedure is analogous
to the analytical continuation of Malikov-Feigin-Fuchs for affine Lie
algebras \cite{MFF}, for which complex exponents of the generators are used.
We disagree with this view because the complex exponents for
the affine Lie algebras, used also by Kent for the Virasoro algebra
\cite{Kent}, involve only well defined generators of the algebra under study.
The intertwiners $g(a,b)$ and $q(a,b)$, however, are made out of a continuum
of `generators' of types $\cG_{\alpha}$ and $\cQ_{\alpha}$, with $\alpha$
a complex number, that do not belong to the Topological N=2 algebra
\req{topalgebra} under study, except for $\a$ integer,  
mixing in fact a continuum of different N=2 algebras. Furthermore,
the authors {\it postulate} a number of properties and results for 
$g(a,b)$ and $q(a,b)$ that they call a `consistent' set of algebraic rules.
In this set, however, important commutators are absent: 
$[g(a,b), \cQ_{\alpha}]$, $[q(a,b), \cG_{\alpha}]$ and $[g(a,b), q(a',b')]$.

Finally, underlying the construction of the `non-conventional' \svec s and 
the idea that they generate `maximal' submodules, is the authors claim
\cite{FST}\cite{Seiso}\cite{SeSi} that the `topological' and `massive' 
non-conventional \svec s of `the N=2 algebra' (see footnote 1) are in one-to-one 
correspondence with the conventional and `relaxed' \svec s of the affine 
$\hat sl(2)$ algebra, respectively. The `relaxed' h.w. vectors and \svec s 
have neither clear meaning nor intrinsic interest from the point of 
view of the affine $\hat sl(2)$ algebra.
They are not annihilated by any of the zero modes $J_0$, unlike the
conventional \svec s, and they satisfy a very strong constraint (apparently
in order to mimic some fermionic properties of `the N=2 algebra'). 
They have been introduced by the authors with the unique purpose to
create an isomorphism between the 
Verma modules and submodules of `the N=2 algebra'
and the Verma modules and submodules of the affine $\hat sl(2)$ algebra.
The authors claim that the `topological' and `massive' submodules generated
by the `topological' and `massive' \svec s of `the N=2 algebra' are 
isomorphic to the conventional and `relaxed' submodules generated by the
conventional and `relaxed' \svec s of the affine $\hat sl(2)$ algebra,
respectively\footnote{The precise claim is that the
`twisted topological' and `twisted massive' \svec s and submodules 
of `the N=2 algebra' are isomorphic to the `twisted conventional' and 
`twisted relaxed' \svec s and submodules of 
the affine $\hat sl(2)$ algebra, respectively, where twisted refers to
the corresponding spectral flows. For convenience we have dropped the word 
`twisted' in all this paragraph.}.
The crucial point now is that the latter must be maximal
submodules since in the affine $\hat sl(2)$ algebra subsingular vectors 
are not supposed to exist. As a consequence, the submodules generated by
the `topological' and `massive' 
non-conventional \svec s of `the N=2 algebra' must be also
maximal. But we have proved in the previous subsection, using a 
counterexample, that these submodules are not maximal. Therefore there are
only two alternatives: either the isomorphism N=2 $\leftrightarrow \hat sl(2)$
proposed by the authors fails or there exist subsingular vectors in the Verma-like
modules (conventional and `relaxed' Verma modules) of the affine $\hat sl(2)$
algebra. We believe that the isomorphism fails for two reasons. First, as we
pointed out in the paragraphs above, the N=2 non-conventional \svec s, which
are the N=2 counterpart of the isomorphism, are objects of very unclear nature.
In our opinion, it is not even clear that they belong to any Verma modules 
and/or submodules of the Topological N=2 algebra, as claimed by the authors. 
Second, the isomorphism already fails, 
in its present form, because in the `relaxed' Verma modules, which are 
the $\hat sl(2)$ partners of the generic `massive' N=2 Verma modules, 
no submodules have been found that could be the partners of the 
`chiral-chiral' or of the `no-label' N=2 submodules. As a matter of fact, it
seems a highly non-trivial task to find the $\hat sl(2)$ partners of the
no-label \svec s (that in the notation used by the authors should be
called no-label subsingular vectors instead, as we have explained before).
We think that these partners simply do not exist.

\subsection{The N=2 embedding diagrams}\lvm

The same misleading ideas used by Semikhatov and Tipunin to obtain
the degeneration patterns and classification of subsingular vectors in
ref. \cite{SeTi3}, were used afterwards by Semikhatov and Sirota in
ref. \cite{SeSi}, where they presented a prescription to obtain a
classification of N=2 embedding diagrams. That is, they applied again 
the assumption that in N=2 Verma modules there are submodules of 
exactly two different types -- the (twisted) `massive' and the (twisted) 
`topological' ones and arbitrary sums thereof -- and the assumption 
that these submodules are generated maximally by the (twisted) `massive' 
and the (twisted) `topological' non-conventional \svec s. Hence it seems 
that there is no need to add any more comments to this issue. Nevertheless, 
there are two important remarks to be added. 

First, the authors did not present a single N=2 embedding diagram in 
\cite{SeSi} (in spite of their claims). Instead they presented a classification of 
embedding diagrams of the $\hat sl(2)$ Verma-like structures, for
which there are only two types of submodules and these are generated
maximally by their corresponding \svec s. Then the authors proposed the
isomorphism that we have discussed in the previous subsection, 
between the N=2 Verma modules and submodules and
the $\hat sl(2)$ Verma-like modules and submodules, and finally the
reader was supposed to apply this isomorphism to obtain, after working 
for several weeks, the sought N=2 embedding diagrams. (In the abstract
of the paper, the first sentence reads however: `We classify and 
explicitly construct the embedding diagrams of Verma modules over the
N=2 supersymmetric extension of the Virasoro algebra').    

Second, the authors did not compare their `classification' of N=2 
embedding diagrams with the most complete classification done so far
on this issue, due to D\"orrzapf in 1995 \cite{Doerrthesis} (Ph. D. thesis,
Cambridge). The authors did not even mention this important work of
D\"orrzapf, abundantly referred, however, in some references cited by
the authors\footnote{This is most intriguing because the classification of N=2 
embedding diagrams in the thesis of D\"orrzapf was well known to Semikhatov 
as it was shown to him in Durham during the academic year 1995-96 and 
he even borrowed the thesis for a few days. In addition, Semikhatov sent
an e-mail to D\"orrzapf in December 1997 where he admitted to know the
work on N=2 embedding diagrams in his thesis \cite{Doerrcom}. 
In spite of these facts Semikhatov has kept ignoring the 
classification of N=2 embedding diagrams by D\"orrzapf also in the papers
\cite{FSST}, \cite{SeTi2} and \cite{FS} that followed ref. \cite{SeSi}.}.

\section{Final Remarks}\lvm

In January 97 our paper \cite{BJI6} appeared in hep-th. As was indicated
in the abstract, all the \svec s at level 1 of the Topological N=2 algebra
were presented, as well as the subsingular vectors which become singular
in chiral Verma modules at levels 2 and 3. Semikhatov and Tipunin only
needed to look (literally!) at the \svec s at level 1 to realize straightforwardly
that their classification of submodules was incomplete as there is no place
left for `no-label' \svec s neither for `chiral-chiral' pairs of uncharged \svec s. 
Moreover, the authors only needed to check the few examples of subsingular 
vectors, which were the only examples known in the literature, to find 
out that their classification of subsingular vectors was incorrect and their claim
that the `non-conventional' \svec s generate maximal submodules was also
incorrect. These incorrect claims they published however three months
after, in April 97 in ref. \cite{SeTi3}. Furthermore in August 97 they
sent to hep-th a revised version of that paper, accepted for publication 
in Comm. Math. Phys., in which they added several misleading claims
about `our' subsingular vectors in ref. \cite{BJI6}. The most disturbing
claim was that they were subsingular (instead of singular) in the chiral
Verma modules.

Because of these incorrect claims we contacted the editors of Comm. Math. 
Phys. who very kindly stopped the publication of the paper \cite{SeTi3} in order
for the authors to revise the paper. We also pointed out, among several
other remarks, that: i) All the results presented in \cite{SeTi3} were based
on the assumption that there are only two different types of submodules in 
N=2 Verma modules, without giving any proof for this strong claim, 
ii) Our classification of h.w. vectors was completely overlooked, in particular 
no-label h.w. vectors were ignored, and iii) Our subsingular vectors, which
are singular in chiral Verma modules, did not fit into 
the classification of subsingular vectors given by the authors.

The authors replied to the editors that they had in fact misclassified the
subsingular vectors in ref. \cite{BJI6}, but however they were described 
by Proposition 3.9 case $n=0$. (In this note we have proved that this is 
not the case). They also assured that they did not take into account
no-label \svec s because they do not exist in generic (`massive') Verma
modules (in ref. \cite{BJI6} we had proved, however, that no-label \svec s
{\it only} exist in generic Verma modules and one can see this also in the
explicit examples at level 1) . They also assured that the spectral flows
transform not only the states but also the h.w. conditions in such a way
that the h.w. vectors are always mapped into h.w. vectors, for any
spectral flow transformation (they also assured that the Verma modules
of the Neveu-Schwarz  N=2 algebra are isomorphic to the Verma modules 
of the Ramond N=2 algebra for this reason). In spite of this misleading reply 
the paper was finally published, although with several improvements. 

In November 98 we explained the content of this note to the authors, in
particular we gave them the proof, step by step, that their `topological'
and `massive' non-conventional \svec s do not generate maximal submodules.
In spite of this the authors (who never replied to us)
still published, in Nucl. Phys. B, the paper \cite{FSST} 
with the usual assumptions that there are only two types of submodules,... For 
example one can read in the second paragraph below eq. (3.6) the statement:
`...these \svec s (the topological ones) generate maximal submodules, which
is crucial for the resolution to have precisely the form (3.2) ...'.  

Finally we would like to point out that the facts discussed in this note affect
drastically the results presented in refs. \cite{SeTi3} -- \cite{FSST} and less
importantly also the results presented in ref. \cite{FS}.

\vskip .3in

{\bf Acknowledgements}\lvm

\small

We thank B.L. Feigin and A. Taormina for encouraging us to 
write this letter and make public our disagreements with the work of
Semikhatov and collaborators about the N=2 superconformal algebras. 
We would like to thank B.L. Feigin also for elucidating to us his involvement 
in the papers \cite{FST}, \cite{FSST} and \cite{FS} and his commitment
to the results. This clarifies the puzzling lack of rigour observed by us, 
and by several other colleagues, in these papers. Finally, we are
indebted to  A. Jaffe and T. Miwa, as editors of Comm. Math. Phys. for their
help in removing the misleading claims about `our subsingular vectors'
from the final version of the paper \cite{SeTi3}, by A. Semikhatov and
Y. Tipunin, published in CMP.

\end{document}